\documentclass{article}

\usepackage[utf8]{inputenc} 

\usepackage{authblk}
\usepackage{fullpage}
\usepackage{subcaption}

\usepackage{geometry} 
\usepackage{graphicx} 

\usepackage{booktabs} 
\usepackage{array} 
\usepackage{paralist} 
\usepackage{verbatim} 
\usepackage{amssymb, amsmath, amsthm, mathabx}
\usepackage{epstopdf}

\usepackage{algorithm2e}

\usepackage{fancyhdr} 
\usepackage{sectsty}
\allsectionsfont{\sffamily\mdseries\upshape} 

\usepackage[nottoc,notlof,notlot]{tocbibind} 
\usepackage[titles,subfigure]{tocloft}
\usepackage{mathrsfs}
\usepackage{bm}
\usepackage{color}
\usepackage{bbm}

\usepackage{cite}
\usepackage{amsmath,amssymb,amsfonts,graphicx}
\usepackage{bm}
\usepackage[nottoc,notlof,notlot]{tocbibind}
\usepackage{animate}

\newcommand{\dd}{\mathrm{ d}}

\newcommand{\atantwo}{\mathrm{arctan2}}
\newcommand{\tr}{\Tilde{r}}

\newtheorem{theorem}{Theorem}[section]

\newtheorem{conjecture*}{Conjecture}

\graphicspath{{./figs/}}

\title{Lightfield Coordinates Adapted to Asgeirsson's Theorem}
\author{Haotian Li, He Qin, Todor Georgiev}
\date{Adobe Systems, San Jose, CA 95110, USA}

\begin{document}

\maketitle

\section*{Abstract}
John's differential equation and its canonical form, the ultrahyperbolic equation, plays important role in lightfield imaging. The equation describes a local constraint on the lightfield, that was first observed as a ``dimensionality gap'' \cite{5539854} in the frequency representation. Related to the ultrahyperbolic equation, Asgeirsson's theorems describe global properties. These indicate new, global, constraints on the lightfield. 
In order to help validate those theorems on real captured images, we introduce a coordinate system for the lightfield, which suits better the Asgeirsson theorems, and analyze behaviour in terms of the new coordinates.

\noindent\textbf{Keywords:} Lightfield, John's equation, ultrahyperbolic PDE, Asgeirsson's theorems, $4D$ radiance.

\section{Introduction}

\subsection{John's Transform and John's Equation}
\begin{figure}[hbt!]
    \centering
    \includegraphics[width = 0.4\textwidth]{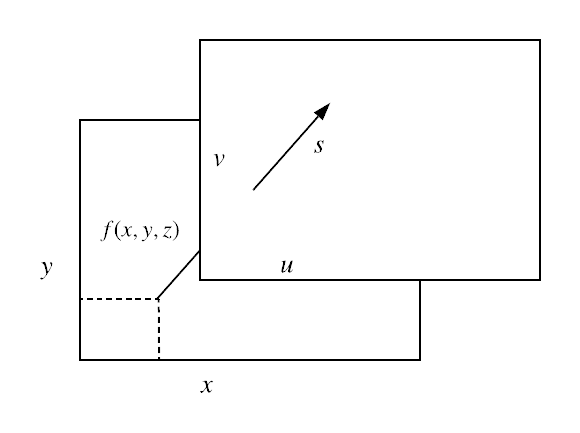}
    \caption{Two parallel planes parameterization of straight lines in $3D$.}
    \label{fig: JT}
\end{figure}

Given a function $f$ describing density of isotropic light sources in $3D$, the John transform $r$ of $f$ is defined as its integral along any straight line $\xi$:
\begin{align}
    J(f) = r, \qquad r(\xi) := \int_{\xi} f(x,y,z) \dd m(x,y,z),
\end{align}
where $\dd m$ is the Euclidean measure on the straight lines $\xi$ (notations taken from~\cite{helgason1999radon}). If we use two-plane parametrization for $\xi$ (Fig.~\ref{fig: JT}), where $(x,y)$ gives the intersection of a light ray $\xi$ with the first plane and $(u,v)$ indicates the angles by tracking the displacements of $\xi$ on the second plane, then
\begin{align}
    J(f) = r, \qquad r(x,y,u,v) = \int_{-\infty}^\infty f(x + u z, y + v z, z) dz
    \label{eqn: John Transform}
\end{align}

John's equation is derived from Eq.~(\ref{eqn: John Transform}). It constrains the radiance $r$ in the following way:
\begin{align}
    (\frac{\partial}{\partial y} \frac{\partial}{\partial u} - \frac{\partial}{\partial x} \frac{\partial}{\partial v}) r = 0
    \label{eqn: John's equation}
\end{align}

\subsection{The Ultrahyperbolic Equation}
Furthermore, if we do the following reparametrization of $r(x,y,u,v)$ into $\tr(\xi_1,\xi_2,\xi_3,\xi_4)$ (i.e. transform from $4D$ $(x,y,u,v)$-space to $4D$ $(\xi_1,\xi_2,\xi_3,\xi_4)$-space),
\begin{align}
    \begin{cases}
    \xi_1 = \frac{1}{2}(u + y)\\
    \xi_2 = \frac{1}{2}(u - y)\\
    \xi_3 = \frac{1}{2}(v + x)\\
    \xi_4 = \frac{1}{2}(v - x)
    \end{cases}
    \qquad \qquad
    \begin{cases}
    x = \xi_3 - \xi_4\\
    y = \xi_1 - \xi_2\\
    u = \xi_1 + \xi_2\\
    v = \xi_3 + \xi_4
    \end{cases}
    \label{eqn: linear transform}
\end{align}
we get the ultrahyperbolic partial differential equation
\begin{align}
    (\partial_{\xi_1\xi_1} - \partial_{\xi_2\xi_2} - \partial_{\xi_3 \xi_3} + \partial_{\xi_4 \xi_4})\tr =0 \qquad \text{ or equivalently, } (\Delta_{14} - \Delta_{23}) \tr = 0,
    \label{eqn: ultrahyperbolic}
\end{align}
where $\Delta_{14}$ and $\Delta_{23}$ are the Laplacians in the $(\xi_1,\xi_4)$ and $(\xi_2,\xi_3)$ planes, respectively. (See~\cite{georgiev2019john}).

\subsection{Asgeirsson's Theorems}
Next we use the ultrahyperbolic equation with four variables as shown in Eq.~\eqref{eqn: ultrahyperbolic}. We base our analysis on the following theorems by Asgeirsson (see \cite{courant2008methods}). 

\begin{theorem}
Integral over a circle $C_1$ with radius $R$ in the $(1,4)$-plane is equal to the integral over the same radius circle $C_2$ in the $(2,3)$-plane, i.e.,
\begin{align}
     \int_{0}^{2\pi} r(\xi_1 + R \cos \theta,\xi_2,\xi_3,\xi_4 + R \sin \theta) \dd \theta = \int_{0}^{2\pi} r(\xi_1, \xi_2 + R \cos \theta,\xi_3+ R \sin \theta, \xi_4) \dd \theta
\end{align}
\end{theorem}

\begin{theorem}
More generally, if we consider a double integral over two circles, one of which has radius $R_1$ in $(1,4)$-plane and the other has radius $R_2$ in $(2,3)$-plane, it is equal to the double integral over two circles with two radii switched in the two planes, i.e.,
\begin{align*}
     &\int_0^{2\pi} \int_{0}^{2\pi} r(\xi_1 + R_1 \cos \theta_1,\xi_2 + R_2 \cos \theta_2,\xi_3 + R_2 \sin \theta_2,\xi_4 + R_1 \sin \theta_1) \dd \theta_1 \dd \theta_2 \\&= \int_{0}^{2\pi} \int_{0}^{2\pi} r(\xi_1 + R_2 \cos \theta_1,\xi_2 + R_1 \cos \theta_2,\xi_3 + R_1 \sin \theta_2,\xi_4 + R_2 \sin \theta_1) \dd \theta_1 \dd \theta_2
\end{align*}
\label{thm: Asgeirsson}
\end{theorem}

\section{New Coordinates}
We introduce two sets of polar coordinates in the planes $(\xi_1,\xi_4)$ and $(\xi_2,\xi_3)$, respectively:
\begin{align*}
    \begin{cases}
    \xi_1 = R_1 \cos \theta_1 \\
    \xi_4 = R_1 \sin \theta_1
    \end{cases}
    \qquad 
    \begin{cases}
    \xi_2 = R_2 \cos \theta_2 \\
    \xi_3 = R_2 \sin \theta_2 
    \end{cases}
    \qquad
    \text{where } R_1, R_2 \geq 0 \text{ and } \theta_1,\theta_2 \in [0,2\pi).
\end{align*}
Then, we consider the new coordinate system $(\theta_1, \theta_2, R_1, R_2)$ for the lightfield, in which $(\theta_1, \theta_2)$ are the coordinates within each new microimage, and $(R_1, R_2)$ are the coordinates of the new microimages. In this coordinate system Theorem~\ref{thm: Asgeirsson} states that the double integral over ${(\theta_1,\theta_2)}$ of a microimage with coordinates $(R_1, R_2)$ is equal to the double integral over ${(\theta_1,\theta_2)}$ of a microimage with coordinates $(R_2, R_1)$.

\section{Change of Coordinates}

\textbf{Coordinate Transformation:} $(\theta_1,\theta_2,R_1,R_2)\longrightarrow(x,y,u,v)$
\begin{align}
    \begin{cases}
    & x = R_2 \sin \theta_2 - R_1 \sin \theta_1 \\
    & y = R_1 \cos \theta_1 - R_2 \cos \theta_2 \\
    & u = R_1 \cos \theta_1 + R_2 \cos \theta_2 \\
    & v = R_2 \sin \theta_2 + R_1 \sin \theta_1 \\
    \end{cases}
    \label{eqn: coord. transformation}
\end{align}

\begin{figure}
    \centering
    \includegraphics[width = 0.5\textwidth]{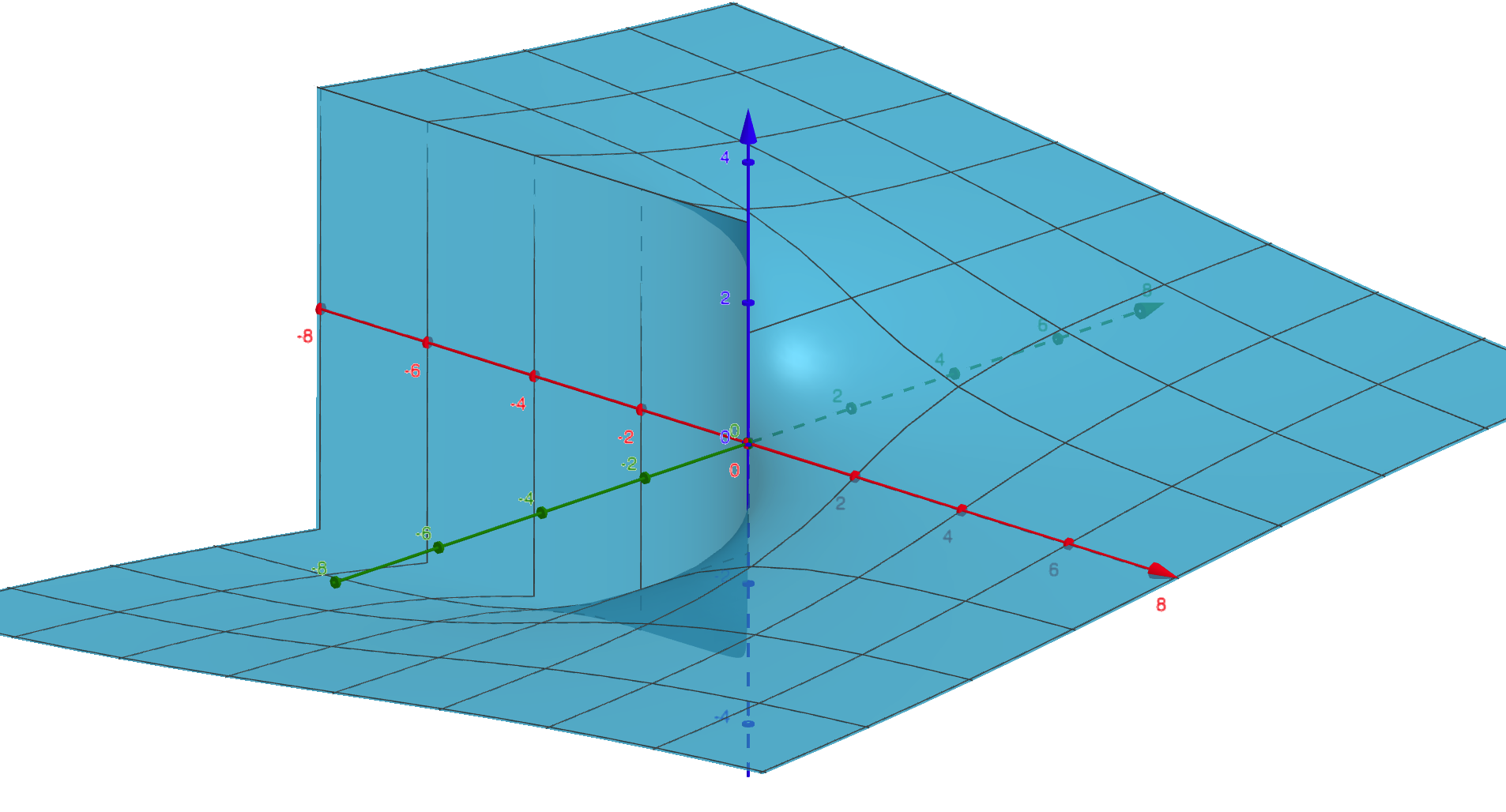}
    \caption{Function $z = \atantwo(y,x)$, $x$-axis (green), $y$-axis (blue), $z$-axis (red)}
    \label{fig:arctan2}
\end{figure}

\bigskip
\noindent \textbf{Inverse Coordinate Transformation:} $(x,y,u,v) \longrightarrow (\theta_1,\theta_2,R_1,R_2)$

\begin{align}
    \begin{cases}
    & \theta_1 = \atantwo (v-x, u+y) \\
    & \theta_2 = \atantwo (v+x, u-y) \\
    & R_1 = \frac{1}{2}\sqrt{(u+y)^2 + (v-x)^2} \\
    & R_2 = \frac{1}{2}\sqrt{(u-y)^2 + (v+x)^2} \\
    \end{cases}
\end{align}
where $\atantwo$ is a variation form of $\arctan$ (see Fig.~\ref{fig:arctan2}):
\begin{align}
    \atantwo(y,x) = 
    \begin{cases}
        \arctan(\frac{y}{x}) & \text{if }x>0,\\
        \arctan(\frac{y}{x}) + \pi & \text{if }x<0 \text{ and } y \geq 0, \\
        \arctan(\frac{y}{x}) - \pi & \text{if }x<0 \text{ and } y < 0, \\
        \frac{\pi}{2} & \text{if }x=0 \text{ and } y > 0, \\
        -\frac{\pi}{2} & \text{if }x=0 \text{ and } y < 0, \\
        \text{undefined} & \text{if }x=0 \text{ and } y = 0. \\
    \end{cases}
\end{align}

\section{Discretization}
\begin{figure}
    \centering
    \includegraphics[width = .4\textwidth]{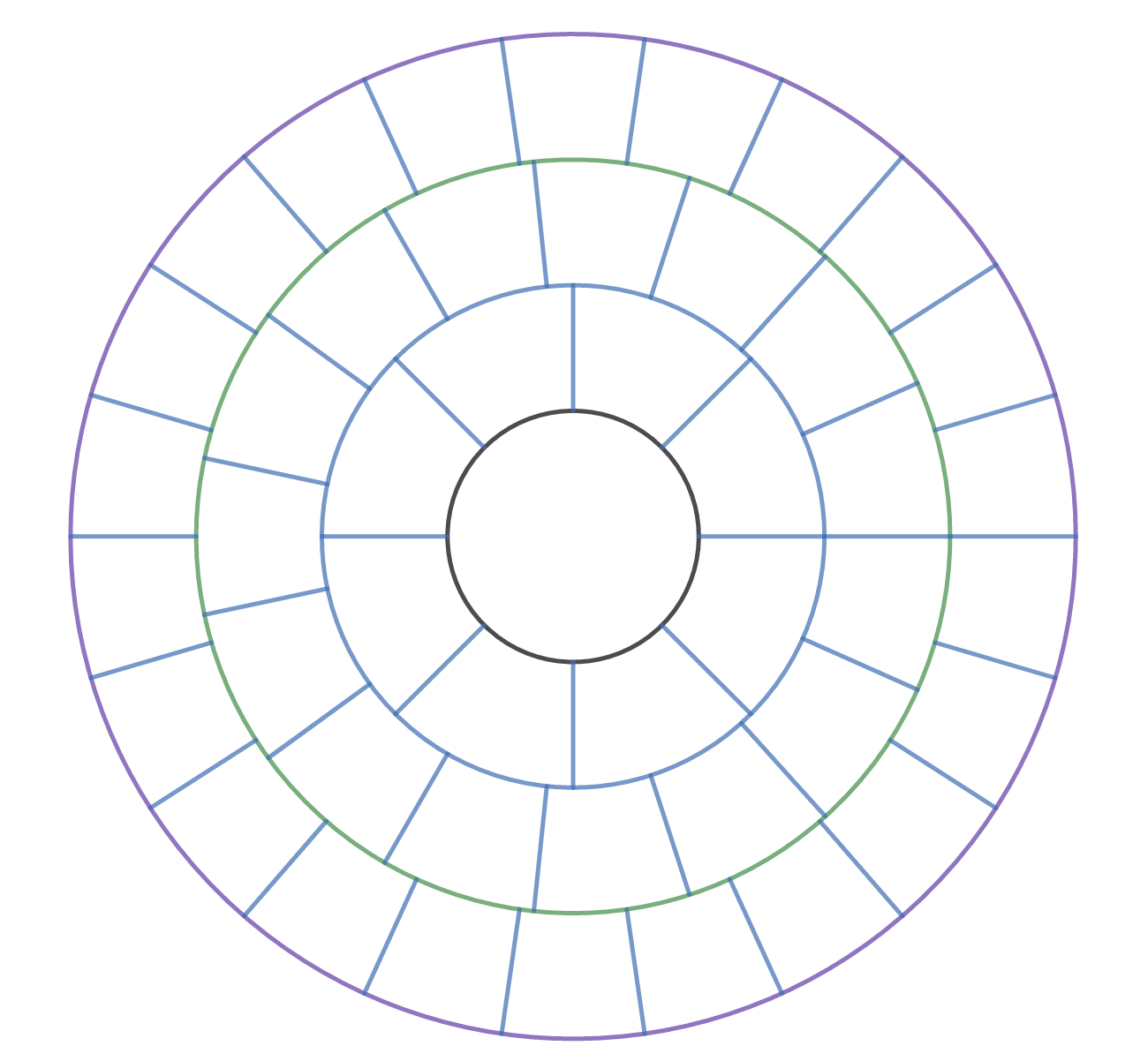}
    \caption{$R_1, \theta_1$ discretization in plane $(\xi_1,\xi_4)$. The concentric rings represent $R_1 = 0,1,2,3.$}
    \label{fig:polar_discretization}
\end{figure}

To deal with real images with pixels, discretization of the previous formulas is necessary. Here we discretize the radii and the angles.
First, we simply discretize both radii $R_i,$ $i = 1, 2$, into non-negative integers. For example $R_1 = 0,1,2,3,\cdots$, and similar for $R_2$.
Second, we discretize the angles $\theta_i$, $i = 1, 2$, in a linear (in $R_i$) way: For $\theta_i$, we evenly split the interval $[0,2\pi)$ into $7R_i+1$ sub-intervals (as shown in Fig.~\ref{fig:polar_discretization}). We discretize in a linear fashion in order to produce uniform representation (see below). The Jacobian in polar coordinate is linear w.r.t. the radius value.

\begin{figure}
    \centering
    \includegraphics[width = 0.4\textwidth]{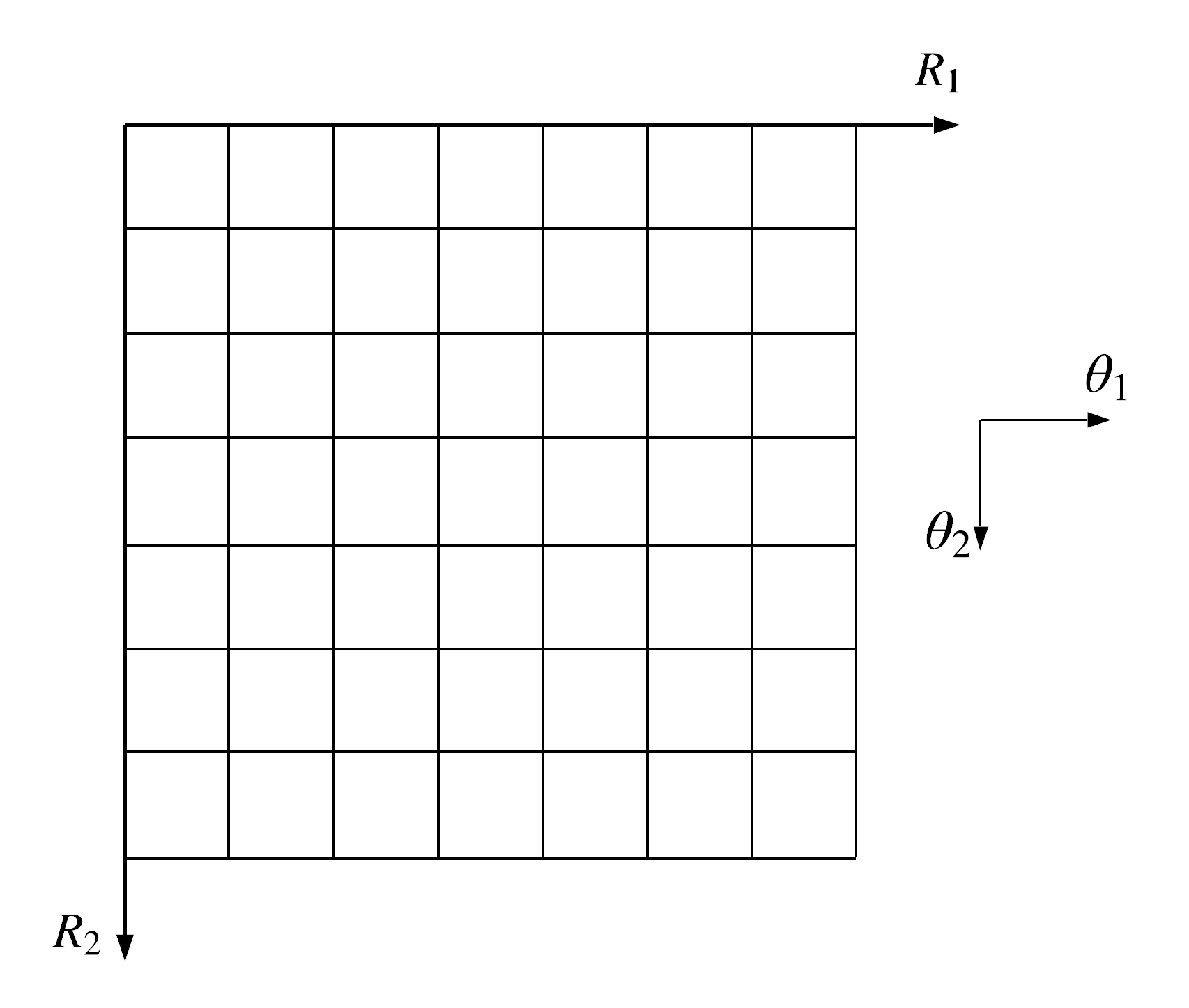}
    \caption{New coordinates for the lightfield. $(\theta_1,\theta_2)$ are the coordinates of pixels inside each of the small squares / rectangles.}
    \label{fig:new coord}
\end{figure}

After discretization, we arrange the coordinates $(\theta_1,\theta_2,R_1,R_2)$ as shown in Fig.~\ref{fig:new coord}. 
Since pixels are equally spaced, the sizes of different microimages are different. The size of a microimage depends on the number of pixels in it.

Now Asgeirsson's theorems simply say that the sum of all pixels in a microimage with coordinates $(R_1, R_2)$ is equal to the sum of all pixels in the corresponding microimage with coordinates $(R_2, R_1)$. Note that those two microimages are symmetric with respect to the main diagonal in Fig.~\ref{fig:new coord}.

\section{Numerical Experiments}
We use the Seagull lightfield image (Fig.~\ref{fig:seagull lightfield}) to generate an example of the new lightfield representation (Fig.~\ref{fig:new coord seagull}). In the experiments, we choose the pixel at the center of the seagull's eye as the reference point, i.e., the origin of the $4D$ coordinates, in the Seagull lightfield.

Since the coordinate transformation (Eq.~\eqref{eqn: coord. transformation}) is continuous and analytic, we generally get floating point number coordinates instead of integers when we try to fill in the new coordinate system's pixel values. So in Fig.~\ref{fig:new coord seagull}, we use nearest neighbor sampling to read the pixel value from the original lightfield image.

Figure~\ref{fig:color mapping of new coord seagull} is used as a map representing the coordinates in Figure 6.

Figure~\ref{fig:theorems} is a zoom in into Figure~\ref{fig:new coord seagull}d, where $R_1$ and $R_2$ take values from $0$ to $4$. 

The first Asgeirsson theorem says that the sum of all pixels with given $R_2 = k$ in the single row at the top, is equal to the sum of all pixels in the single column on the left having $R_1$ equal to the same $k$.

More generally, the second Asgeirsson theorem says that in symmetric boxes pixels sum up to the same value. Boxes are defined symmetric relative to the main diagonal (see Section 4).

\section{Note on Using Shift for Lightfield Imaging}
One important practical point about lightfield imaging is sparsity of sampling, and the resulting aliasing. Due to the spatio-angular resolution tradeoff~\cite{Spatio-Angular2006} and the need to produce higher spatial resolution with limited sensor size, we often perform sparse optical sampling in the angular dimensions. This results in aliasing artifacts when we render the final image, or when we apply any 4D filter.

\begin{figure}
    \centering
    \includegraphics[width = \textwidth]{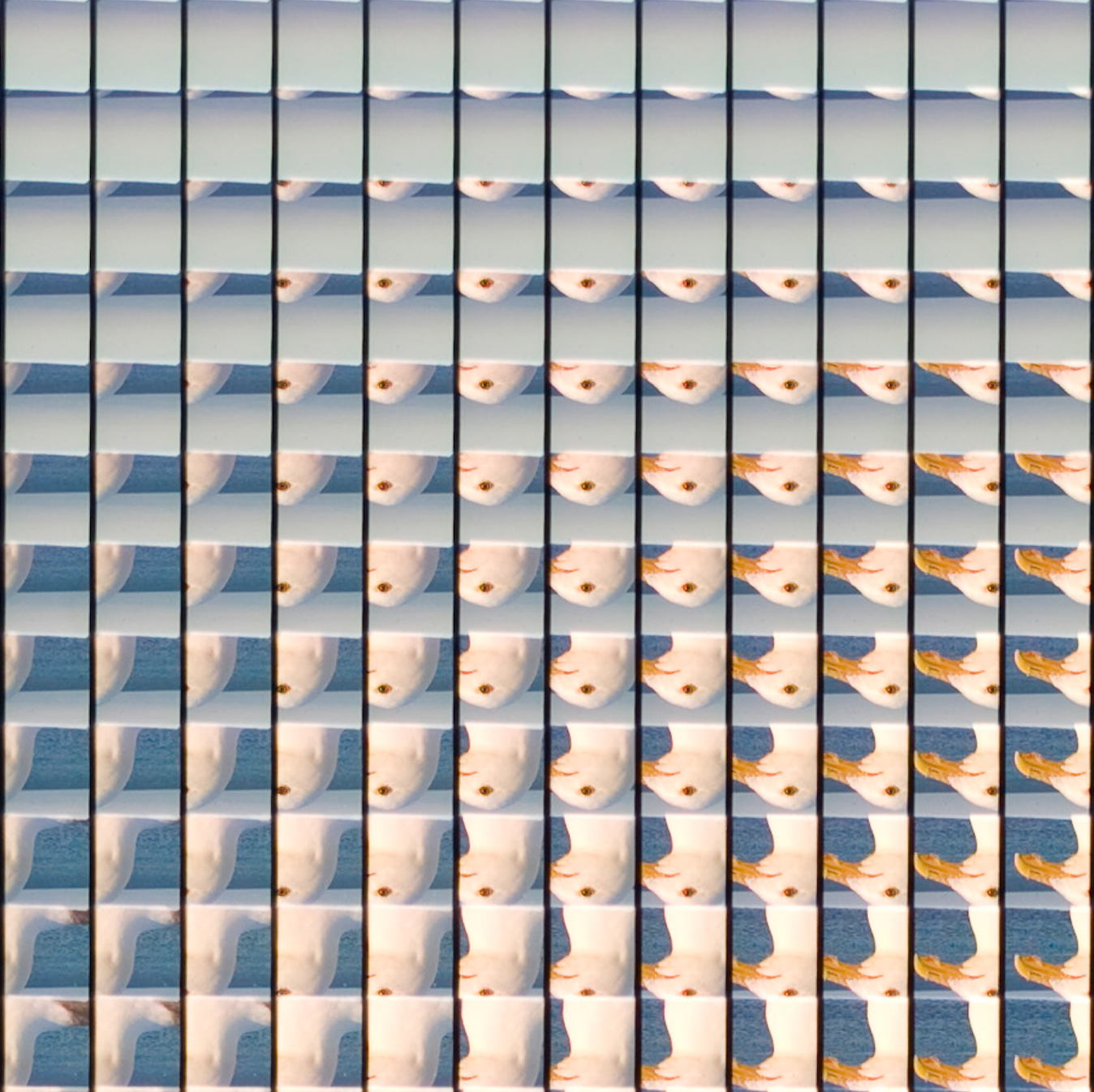}
    \caption{Original lightfield Seagull captured with our Plenoptic 2.0 camera. Coordinates are chosen in the traditional way: Microimage locatedion is parametrized by $(u, v)$, pixels microimage having coordinates $(x, y)$.}
    \label{fig:seagull lightfield}
\end{figure}

\begin{figure}
    \centering
    \begin{subfigure}[b]{0.45\textwidth}
    \centering
    \includegraphics[width = 0.8\textwidth]{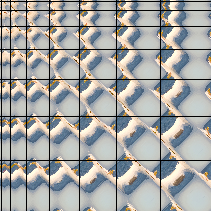}
    \caption{shift = 0}
    \end{subfigure}
    \begin{subfigure}[b]{0.45\textwidth}
    \centering
    \includegraphics[width = 0.8\textwidth]{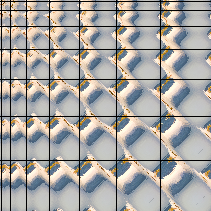}
    \caption{shift = 2}
    \end{subfigure}
    \begin{subfigure}[b]{0.45\textwidth}
    \centering
    \includegraphics[width = 0.8\textwidth]{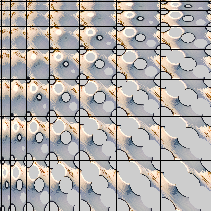}
    \caption{shift = 5}
    \end{subfigure}
    \begin{subfigure}[b]{0.45\textwidth}
    \centering
    \includegraphics[width = 0.8\textwidth]{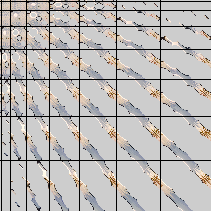}
    \caption{shift = 9}
    \end{subfigure}
    \caption{New coordinate representations of lightfield Seagull (Fig.~\ref{fig:seagull lightfield}) with different shifts. The important role of shifts is explained in section 6. See Fig.~\ref{fig:color mapping of new coord seagull} for colormap and description of our coordinates.}
    \label{fig:new coord seagull}
\end{figure}

\begin{figure}
    \centering
    \begin{subfigure}[b]{0.45\textwidth}
    \centering
    \includegraphics[width = 0.8\textwidth]{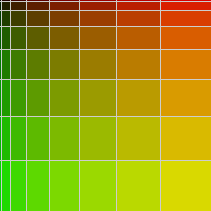}
    \caption{shift = 0}
    \end{subfigure}
    \begin{subfigure}[b]{0.45\textwidth}
    \centering
    \includegraphics[width = 0.8\textwidth]{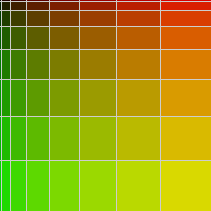}
    \caption{shift = 2}
    \end{subfigure}
    \begin{subfigure}[b]{0.45\textwidth}
    \centering
    \includegraphics[width = 0.8\textwidth]{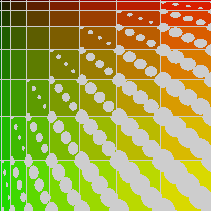}
    \caption{shift = 5}
    \end{subfigure}
    \begin{subfigure}[b]{0.45\textwidth}
    \centering
    \includegraphics[width = 0.8\textwidth]{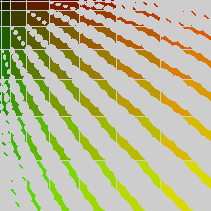}
    \caption{shift = 9}
    \end{subfigure}
    \caption{Color mapping of the new coordinate representation of lightfield Seagull (Fig.~\ref{fig:new coord seagull}) with different shifts. Different tones of green represent values of $R_1$ from $0$ (darkest) to $7$ (lightest), and similar for red (representing $R_2$). Pixels inside each rectangle have coordinates $\theta_1$ and $\theta_2$ ranging from $0$ to $2\pi$. Gray represents pixels that are outside the captured lightfield (in our case outside Fig.~\ref{fig:seagull lightfield}).}
    \label{fig:color mapping of new coord seagull}
\end{figure}

\begin{figure}
    \centering
    \includegraphics[width = 0.5\textwidth]{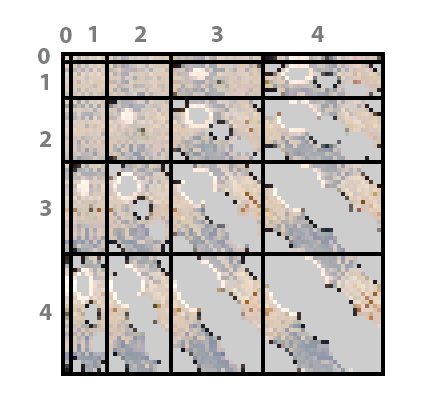}
    \caption{A zoom in into Fig.~\ref{fig:new coord seagull}d, where $R_1$ and $R_2$ take values from $0$ to $4$.}
    \label{fig:theorems}
\end{figure}

This influences our algorithms for rendering of the final image and refocusing. It also influences the computation of derivatives in the angular directions $u$ and $v$. The same pixel location $(x, y)$ in two neighboring microimages $(u, v)$ and $(u+1, v)$ sample two very different rays in the lightfield and their difference cannot represent correctly the derivative of the radiance in $u$-direction. This is due to sparse optical sampling. We need a microimage between $(u, v)$ and $(u+1, v)$, which however has not been captured. Considering epipolar geometry, we substitute the pixel with spatial coordinates $(x, y)$ in this missing microimage with a pixel from the $(u+1, v)$ microimage, with shifted $x$-coordinate. This is the same shift or ``patch size" that is used in lightfield/plenoptic rendering (see for example ~\cite{Focused}). It depends on depth. For the seagull in our lightfield (Fig.~\ref{fig:seagull lightfield}) the shift value is $8.78$ pixels, which we approximate with $9$ in Figures ~\ref{fig:new coord seagull},~\ref{fig:color mapping of new coord seagull},~\ref{fig:theorems}.

\begin{figure}
    \centering
    \includegraphics[width = \textwidth]{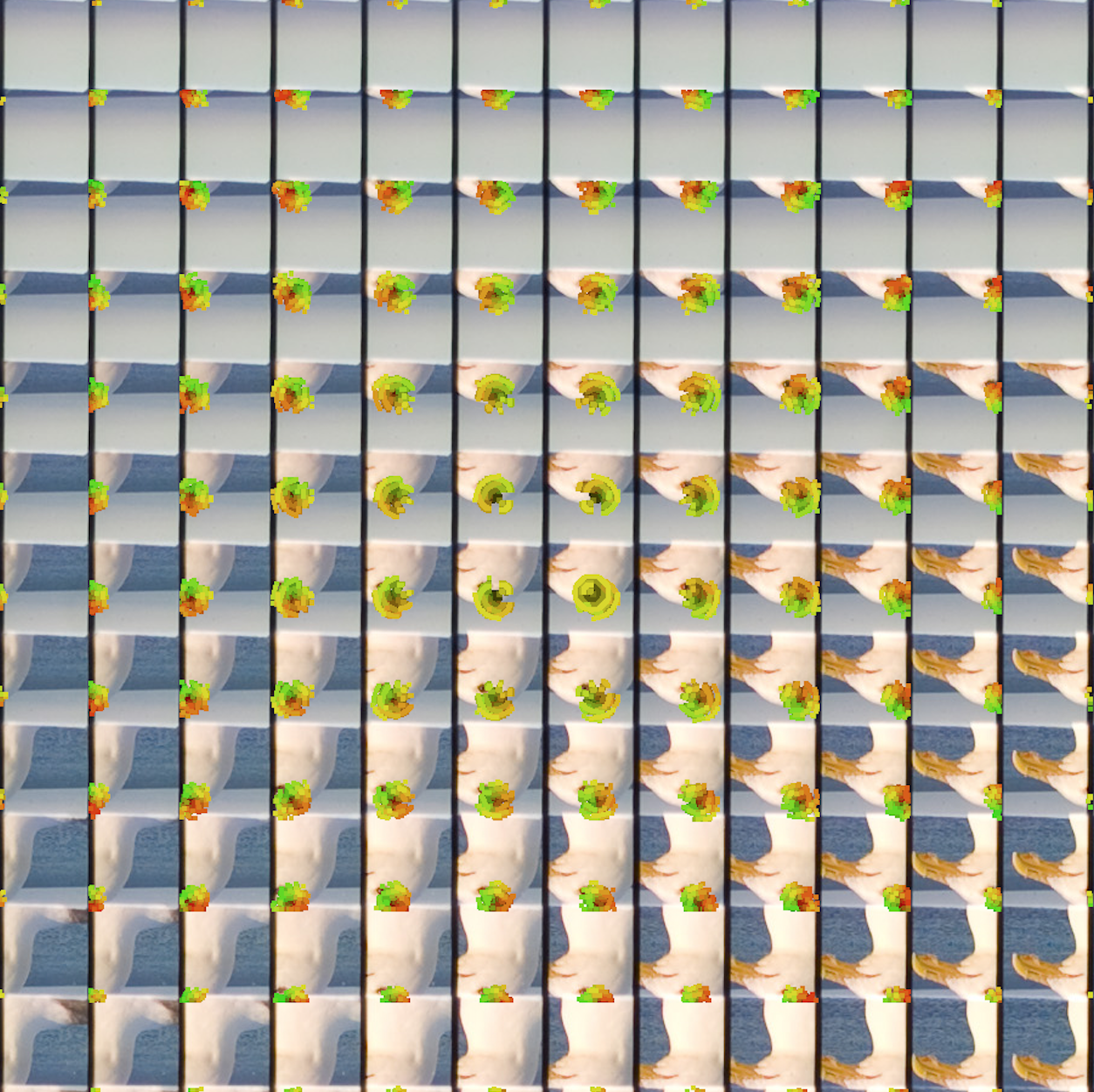}
    \caption{Color mapping in original coordinates with shift 9.}
    \label{fig:mapping in lightfield}
\end{figure}

\clearpage
\bibliographystyle{siam}
\bibliography{ref}

\end{document}